\documentclass[12pt,a4paper]{article}
\usepackage[latin1]{inputenc}
\usepackage{amsmath}
\usepackage{amsfonts}
\usepackage{amssymb}
\usepackage{hyperref}
\usepackage[a4paper]{geometry}
\begin{document}
\author{Levent Akant${^\dag}$, Birses Debir${^\ddag}$, \.I. {\c C}a{\u g}r{\i} \.I{\c s}eri${^\S}$\\
Department of Physics, Bo{\u g}azi{\c c}i University \\ 34342 Bebek, Istanbul, Turkey \\ $^\dag$levent.akant@boun.edu.tr, $^\ddag$birses.debir@boun.edu.tr,
\\${^\S}$islam.iseri@boun.edu.tr}
\title{Non-relativistic Limit of Thermodynamics of Bose Field in a Static Space-time and Bose-Einstein Condensation}\maketitle
\begin{abstract}
We consider the grand canonical thermodynamics of a noninteracting scalar field in a static spacetime. We take the nonrelativistic limit of thermodynamic quantities in a way that leaves the curved structure of the background geometry intact. Using Mellin transform and heat kernel techniques we obtain asymptotic expansions of thermodynamic quantities appropriate for the analysis of Bose-Einstein condensation. We apply our results to investigate gravitational effects on the Bose-Einstein condensation for a scalar field in a finite volume. We also analyze the boundary effects on the depletion coefficient of the scalar field.

\vspace{0.7cm}


\vspace{0.2cm}


\end{abstract}
\newpage
\section{Introduction}

Our aim in this paper is to derive the nonrelativistic limit of the grand canonical thermodynamics of a scalar field in a static spacetime and investigate the gravitational effects on Bose-Einstein condensation in this nonrelativistic limit. The nonrelativistic limit in question is a $c\rightarrow \infty$ limit which is taken in a way that leaves the static background geometry intact. In the ultrarelativistic regime the analysis of the thermodynamics of a quantum field in a static background is well understood \cite{Dowker1, Actor0,Actor,Dowker2,Dowker3,Kirsten2,T1,T2,BECUS}. Here we aim to study the opposite limit.

Taking the nonrelativistic limit of the Klein-Gordon equation on a curved spacetime is a complicated problem. Here we will circumvent this problem by working directly with the free energy. The nonrelativistic limit of the free energy will be taken by applying the saddle point method to an integral expression for the fully relativistic free energy in the limit $c\rightarrow \infty$. An important ingredient of this method is the use of the generalized Laplace method which allows us to keep the background metric intact while making the matter nonrelativistic. In this sense our limit may be regarded as a post-Newtonian approximation to the thermodynamics of the scalar field where the background geometry is treated exactly to all orders in $c$. Although our main focus will be on the static spacetime we will apply our method initially to the special case of an ultrastatic spacetime with the topology of $\mathbf{R}\times \mathbf{M}$ (here $\mathbf{R}$ is the global time). We will show that in this special case our limiting procedure leads, as expected, to the thermodynamics of an ideal Bose gas governed by the Schr\"{o}dinger operator on $\mathbf{M}$. We will then see that the more general static spacetime case can be reduced to the former by a conformal transformation. In the static case the result will involve the Schr\"{o}dinger operator on the optical manifold. It is partly because of these results and partly because of similar usage of the term in the literature \cite{Padmanabhan, Cognola} that we use the term nonrelativistic limit to describe our limiting process.

After deriving the nonrelativistic form of the free energy we will expand the result into an asymptotic series appropriate for the analysis of the system in a large but finite volume, near the critical temperature.
We will make extensive use of Mellin transform and heat kernel techniques in the derivation of this asymptotic series. We will apply our results to examine the gravitational effects on the temperature-density relation, including finite size boundary effects, and on the equation of state. The ultrarelativistic limit will also be considered briefly.

Our starting point will be the fully relativistic functional integral representation of the partition function. As was shown by de Alwis and Ohta in \cite{Alwis} the correct functional measure in a curved spacetime must be constructed with care. The correct measure does not coincide with the measure induced by the natural inner product on the space of field
configurations (the one induced by the kinetic term) and this fact necessitates a conformal transformation which leads to the introduction of the optical metric.
As mentioned above this conformal transformation will also play an important role in the derivation of the $c\rightarrow\infty$ limit of thermodynamics.
The optical metric is an ultrastatic metric if the spacetime metric is static and this fact makes it a convenient tool in the study of quantum field theories on static spacetimes \cite{Dowker1,Gibbons,Fulling,Dowker2,Dowker3,Kirsten2}. However the important connection of the optical metric construction to the functional measure came with \cite{Alwis}. To the best of our knowledge, in the literature on the ultrarelativistic expansion \cite{Dowker3,Kirsten2} the chemical potential has always been introduced after one passes to the optical metric. In contrast here we will introduce the chemical potential in the physical spacetime right at the outset and investigate its fate under the conformal transformation that leads to the optical metric. Since the external fields in a Lagrangian are not usually invariant under a conformal transformation, this is an important question one must address if one aims, as we do here, for a field theoretic derivation of the grand canonical potential. We will show that $\mu$ is invariant under the conformal transformation and thus justify the starting point of the previous works cited above.

It is well known that in the presence of horizons thermodynamic quantities diverge as they are approached \cite{Barbon, Alwis}. In this work we will assume that the system is confined in a region away from the horizon. We plan to study horizon divergences in a future work. We also plan to extend our analysis to the interacting case and to the cosmological backgrounds where the nonrelativistic matter \cite{Lopez, Matos} and nonrelativistic axion dynamics \cite{Sikivie1, Sikivie2} has attracted some recent interest in the context of scalar field dark matter models \cite{sdm}.

Here is the outline of the paper. In Sec. 2 we include the chemical potential into the functional integral for the partition function and generalize de Alwis and Ohta's derivation \cite{Alwis} to this case. The positivity properties of the covariance operator of the functional Gaussian integral representing the partition function of the system are also examined. Then we study some of the properties of the free energy which are important in deriving its nonrelativistic limit. In Sec. 3 we derive $c\rightarrow\infty$ limit of the free energy first in an ultrastatic
spacetime and then, after a careful counting of the factors of $c$ coming from the background metric, we generalize the analysis to the static case. In Sec. 4 we derive aforementioned asymptotic expansions of the free energy and of the occupation number by Mellin transform and heat kernel techniques. In Sec. 5 we apply our results to the Bose-Einstein condensation
in a static spacetime. We also discuss the ultrarelativistic case using the well known ultrarelativistic (high temperature) expansion of the free energy \cite{Dowker1,Actor0,Actor,Dowker2,Dowker3,Kirsten2,T2,BECUS}. In the Appendix A we give the details of the computations described in Sec. 2. In Appendix B we give the details of $\zeta$ function calculations leading to results used in Sec 5.3.

\section{Free Energy in a Static Spacetime}

In this section we will generalize de Alwis and Ohta's analysis \cite{Alwis} of the thermodynamics of a Bose field to include the chemical potential $\mu$. We will also investigate the positivity properties of the covariance operator of the Gaussian functional integral representing the partition function, and determine the set of allowed values of the chemical potential.

We will work in a spacetime region with the static metric
\begin{equation}\label{staticmetric}
  ds^{2}=-F(x)dt^{2}+h_{ij}(x)dx^{i}dx^{j},
\end{equation}
 where $F>0$ and $h_{ij}$ is positive definite. We will confine our system in a spatial region (a submanifold of $t=const.$ hypersurface) $B$ with boundary $\partial B$ and assume the field $\phi$ satisfies either the Neumann boundary condition $\left.N^{\mu} \partial_{\mu}\phi\right|_{\partial B}=0$ or the Dirichlet boundary condition $\left.\phi\right|_{\partial B}=0$. Here $N=N^{\mu} \partial_{\mu}$ is the inward looking unit normal vector field to $\partial B$.

\subsection{Incorporating the Chemical Potential}

In order to incorporate the chemical potential we examine the complex scalar field. It will be convenient to work explicitly with the real and imaginary components $\phi_{a}$ ($a=1,2$) of the field $\phi$. Then the Lagrangian density is given as
\begin{equation}
  \mathcal{L}=\sqrt{|g|}\left[\frac{1}{2}\,g^{\mu\nu}\partial_{\mu}\phi_{a}\partial_{\nu}\phi_{a}+\frac{1}{2}V\phi_{a}\phi_{a}\right].
\end{equation}
Here $g_{\mu\nu}$ is the space-time metric (\ref{staticmetric}) (with Lorentzian signature $\{- + + + \})$, $g=\det g_{\mu\nu}$, $|g|=-g$ and
\begin{equation}
  V=m^{2}+\xi R+V_{ext},
\end{equation}
where $R$ is the scalar curvature of $g_{\mu\nu}$, $\xi$ is a coupling constant for the curvature coupling, and $V_{ext}$ is a possible external potential.

The conserved $U(1)$ current is
\begin{equation}
  j^{\mu}=\sqrt{|g|}g^{\mu\nu}(\phi_{2}\partial_{\nu}\phi_{1}-\phi_{1}\partial_{\nu}\phi_{2}),
\end{equation}
and the momenta are given by
\begin{eqnarray}
  \pi_{a}&=&\sqrt{|g|}g^{00}\partial_{0}\phi_{a}=\sqrt{\frac{h}{F}}\partial_{0}\phi_{a}.
\end{eqnarray}

The Hamiltonian density with the chemical potential added is given by
\begin{eqnarray}\label{Hamiltonian}
  \mathcal{H} =
  \frac{1}{2}\sqrt{\frac{F}{h}}\pi_{a}\pi_{a}-\mu(\pi_{1}\phi_{2}-\pi_{2}\phi_{1})+\sqrt{Fh}\left[\frac{1}{2}h^{ij}\partial_{i}\phi_{a}\partial_{j}\phi_{a}+\frac{1}{2}V\phi_{a}\phi_{a}\right],
  \nonumber \\
\end{eqnarray}
and the Hamiltonian is
\begin{equation}
 H=\int d^{3}x \,\mathcal{H}.
\end{equation}
   The statistical mechanical partition function for the quantum field is
\begin{equation}\label{ztan}
  Z(\beta)=\textrm{Tr}\, e^{-\beta \hat{H}}.
\end{equation}
	and the free energy is given as
\begin{equation}\label{ztanF}
  \mathcal{F}(\beta)=-\frac{1}{\beta}\log Z(\beta).
\end{equation}
  Here $\beta=T^{-1}$ is the inverse temperature (we set Boltzmann constant to one $k_{B}=1$), and $\hat{H}$ is the second quantized field Hamiltonian corresponding to (\ref{Hamiltonian}).
  The \textit{phase space} functional integral representation for the partition function is given as \cite{kapusta, Alwis}
\begin{equation} \label{path}
  Z(\beta)=\int \mathcal{D}\pi\mathcal{D}\phi\,e^{-\int_{0}^{\beta}dt\int d^{d}x\left( \mathcal{H}-i\pi_{a}\partial_{0}\phi_{a}\right)}.
\end{equation}
  Here the integration is over fields $\phi_{a}(t,x)$ satisfying $\phi_{a}(0,x)=\phi_{a}(\beta,x)$, and
\begin{equation}
  \mathcal{D}\pi\mathcal{D}\phi=\prod_{t,x}\prod_{a=1}^{2}\,d\pi_{a}(t,x)d\phi_{a}(t,x).
\end{equation}
  Note that since $Z$ is the statistical partition function there is no factor of $i$ multiplying $\mathcal{H}$  in \ref{path}.

Now we have
\begin{eqnarray}\label{hip}
  \mathcal{H}-i\pi_{a}\dot{\phi}_{a} &=&
  \frac{1}{2}\sqrt{\frac{F}{h}}\pi_{1}^{2}+(-\mu\phi_{2}-i\dot{\phi}_{1})\pi_{1}+\frac{1}{2}\sqrt{\frac{F}{h}}\pi_{2}^{2}+(\mu\phi_{1}-i\dot{\phi}_{2})\pi_{2}+\nonumber \\
   &&+\sqrt{Fh}\left[\frac{1}{2}h^{ij}\partial_{i}\phi_{a}\partial_{j}\phi_{a}+\frac{1}{2}V\phi_{a}\phi_{a}\right].
\end{eqnarray}
  The integral over field momenta is Gaussian and can be computed easily, leading to the configuration space functional integral
\begin{eqnarray}
 Z(\beta)=\int \mathcal{D}\phi\,e^{-S_{E}}.
\end{eqnarray}
 Thanks to the factor of $i$ in the $-i\pi_{a}\dot{\phi}_{a}$ term in (\ref{hip}), the integration over field momenta results in the Euclidean action  \cite{kapusta}
\begin{eqnarray}
 S_{E}=\int dtd^{d}x \sqrt{g_{E}}\left[\frac{1}{2}g_{E}^{\mu\nu}\partial_{\mu}\phi_{a}\partial_{\nu}\phi_{a}+\frac{1}{2}\, \phi_{a}\left(V-\mu^{2}F^{-1}\right)\phi_{a}-iF^{-1}\mu(\dot{\phi}_{1}\phi_{2}-\dot{\phi}_{2}\phi_{1})\right],
\end{eqnarray}
with the Riemannian metric
\begin{equation}
   (g_{E})_{\mu\nu}=\left(
                     \begin{array}{cc}
                       F & 0 \\
                       0 & h_{ij} \\
                     \end{array}
                   \right).
\end{equation}
Note that if the fields obey Dirichlet or Neumann boundary conditions in a finite region with boundary then integration by parts gives
\begin{eqnarray}
 S_{E}
=\int dtd^{d}x
  \sqrt{g_{E}}\left[\frac{1}{2}\,\phi_{a}\left(-\Delta+V-\mu^{2}F^{-1}\right)\phi_{a}-iF^{-1}\mu(\dot{\phi}_{1}\phi_{2}-\dot{\phi}_{2}\phi_{1})\right],
\end{eqnarray}
where $\Delta$ is the Laplacian associated to $(g_{E})_{\mu\nu}$.

As observed by de Alwis and Ohta in \cite{Alwis} the integration over momenta also leads to the functional measure
\begin{equation}\label{measure}
 \mathcal{D}\phi= \prod_{t,x}\prod_{a=1}^{2}\left(\frac{g_{E}(x)}{F^{2}(x)}\right)^{1/4}d\phi_{a}(t,x).
\end{equation}
The standard inner product on the space of field configurations is determined by the kinetic term of this action and is given as
\begin{equation}
  \langle \delta\phi|\delta\psi\rangle=\int dtd^{d}x
  \sqrt{g_{E}}\delta_{ab}\delta\phi_{a}\delta\psi_{b}.
\end{equation}
The functional measure corresponding to this inner product is
\begin{equation}
  \prod_{t,x}\prod_{a=1}^{2}\left[g_{E}(x)\right]^{1/4}d\phi_{a}(t,x).
\end{equation}
Obviously this measure does not match with (\ref{measure}), except in an ultrastatic spacetime where $F=1$. As shown in \cite{Alwis} this mismatch has nontrivial effects on the thermodynamics of the quantum field.

Following \cite{Alwis} let us perform the conformal transformation
\begin{equation}
  \overline{\phi}=F^{\frac{d-1}{4}}\phi,
\end{equation}
and
\begin{equation}
 \overline{g}_{\mu\nu}= F^{-1}(g_{E})_{\mu\nu}=\left(
                     \begin{array}{cc}
                       1 & 0 \\
                       0 & \gamma_{ij} \\
                     \end{array}
                   \right),\;\;\;\;\gamma_{ij}=\frac{h_{ij}}{F}.
\end{equation}
Note that the new metric $\overline{g}_{\mu\nu}$, which is the optical metric, is an ultra-static metric.
Also note that
\begin{equation}
  \sqrt{\overline{g}}= \sqrt{\gamma}=F^{-\frac{d+1}{2}}\sqrt{|g|}.
\end{equation}
If the field $\phi$ satisfies Neumann boundary condition, $\overline{\phi}$ satisfies the Robin (generalized Neumann) boundary condition
\begin{equation}\label{robin}
  N^{\mu} \partial_{\mu}\overline{\phi}+\left[\frac{d-1}{4}N^{\mu} \partial_{\mu}(\log F)\right]\overline{\phi}=0
\end{equation}
where $N=N^{\mu} \partial_{\mu}$ is the inward looking unit normal vector field to $\partial B$. On the other hand if $\phi$ satisfies Dirichlet boundary condition so does $\overline{\phi}$.

As a result of the above conformal transformation we get
\begin{eqnarray}\label{Zfunc}
  Z &=& \int\mathcal{D}\overline{\phi_{1}} \mathcal{D}\overline{\phi_{2}}\,e^{-\int dt\,d^{d}x
  \sqrt{\overline{g}}\frac{1}{2}\left[\overline{\phi}_{a}(-\partial_{0}^{2}-\Delta_{\gamma}+U-\mu^{2})\overline{\phi}_{a}+
  -i\mu(\overline{\phi}_{2}\partial_{0}\overline{\phi}_{1}-\overline{\phi}_{1}\partial_{0}\overline{\phi}_{2})\right]}\nonumber\\
   &=&\int\mathcal{D}\overline{\phi_{1}} \mathcal{D}\overline{\phi_{2}}\,e^{-\int dt\,d^{d}x \sqrt{\overline{g}}\frac{1}{2}\overline{\phi}_{a}\mathcal{A}_{ab}\overline{\phi}_{b}}.
\end{eqnarray}
Here the measure is given by
\begin{equation}
  \mathcal{D}\overline{\phi_{a}}= \prod_{t,x}\prod_{a=1}^{2}\left(\overline{g}(x)\right)^{1/4}d\overline{\phi}_{a}(t,x).
\end{equation}
and
\begin{equation}
  \mathcal{A}=\left(
                \begin{array}{cc}
                  -c^{-2}\partial_{0}^{2}+ A-\mu^{2}c^{-2} & 2ic^{-2}\mu\partial_{0} \\
                  -2ic^{-2}\mu\partial_{0} &  -c^{-2}\partial_{0}^{2}+A-\mu^{2}c^{-2} \\
                \end{array}
              \right)
\end{equation}
with
\begin{equation}\label{A0}
  A=-\Delta_{\gamma}+m^{2}c^{2}+U,
\end{equation}
and
\begin{equation}
  U=\frac{d-1}{4d}R_{\gamma}+F\left(\xi-\frac{d-1}{4d}\right)R+(F-1)m^{2}c^{2}+FV_{ext}.
\end{equation}
In the above formulas we wrote the factors of $c$ explicitly.

The Gaussian integral in (\ref{Zfunc}) is well defined only if $-\partial_{0}^{2}+c^{2}A-\mu^2$ is a positive operator. Since $-\partial_{0}^{2}$ is a positive operator with zero eigenvalue the Gaussian integral will be well defined only if $c^{2}A-\mu^2$ is positive. In Section 2.2 we will show that under suitable conditions $c^{2}A$ is a positive operator
whose lowest eigenvalue $\epsilon_{0}^{2}$ satisfies bounds of the form $\xi_{0}+(VF)_{min}\leq \epsilon_{0}^{2}\leq \xi_{0}+ (VF)_{max} $. Then the set of allowed $\mu$ values is $|\mu|<\epsilon_{0}$.

Scaling the fields as $\overline{\phi}_{a}\rightarrow c\overline{\phi}_{a}$ and performing the Gaussian integral we arrive at
\begin{equation}
  Z= \left[\textrm{det}\left(
                \begin{array}{cc}
                  -\partial_{0}^{2}+c^{2}A-\mu^{2} & 2i\mu\partial_{0} \\
                  -2i\mu\partial_{0} & -\partial_{0}^{2}+ c^{2}A-\mu^{2} \\
                \end{array}
              \right)\right]^{-1/2}.
\end{equation}
Since $ -\partial_{0}$ and $c^{2}A$ commute with each other we get, in terms of their respective eigenvalues $\omega_{n}$ (Matsubara frequencies) and $\epsilon_{\sigma}^{2}$, the following expression for $Z$
\begin{eqnarray}
  Z= \left\{\prod_{n}\prod_{\sigma}\left[(\omega_{n}^{2}+\epsilon_{\sigma}^{2}-\mu^{2})^{2}+4\mu^{2}\omega^{2}_{n}\right]^{-1/2}\right\}.
 \end{eqnarray}
Using the factorization \cite{kapusta}
\begin{eqnarray}\label{factor}
(\omega_{n}^{2}+\epsilon_{\sigma}^{2}-\mu^{2})^{2}+4\mu^{2}\omega^{2}_{n}=\left[\omega_{n}^{2}+(\epsilon_{\sigma}-\mu)^{2}\right]\left[\omega_{n}^{2}
  +(\epsilon_{\sigma}+\mu)^{2}\right],
\end{eqnarray}
 we get the free energy
\begin{eqnarray}\label{free00}
  \mathcal{F}&=&-\frac{1}{\beta}\log Z=\frac{1}{2\beta}\sum_{n}\sum_{\sigma}\left\{\log[\omega_{n}^{2}+(\epsilon_{\sigma}-\mu)^{2}]+\log[\omega_{n}^{2}+(\epsilon_{\sigma}+\mu)^{2}]\right\}\nonumber\\
  &=&\frac{1}{2\beta}\sum_{n}\left\{\log\textrm{det}[\omega_{n}^{2}+(c\sqrt{A}-\mu)^{2}]+\log\textrm{det}[\omega_{n}^{2}+(c\sqrt{A}+\mu)^{2}]\right\}
\end{eqnarray}
 Let us remark that alternative factorizations of the quartic term in (\ref{factor}) lead to multiplicative anomalies in the functional determinant \cite{Dowkermult, Filippi, Tomsmult}. However as shown in \cite{Tomsmult} at least in the Minkowski spacetime the factorization (\ref{factor}) is the one which yields the same result as that obtained by the canonical method, and therefore will be the one used in this work.

Now the expression (\ref{free00}) is similar to the one for the Minkowski case \cite{kapusta}. However unlike the Minkowski case here we do not know the spectrum and the density of states of the operator $A$. Nevertheless, an analysis  based on the zeta function techniques (details are given in the Appendix A) yields not only the expected result
\begin{eqnarray}\label{freesum}
  \mathcal{F}&=&\frac{1}{\beta}\left[\textrm{Tr}\log(1-\,e^{-\beta(c\sqrt{A}- \mu)})+\textrm{Tr}\log(1-\,e^{-\beta(c\sqrt{A}+ \mu)})\right]\\
  &=&\frac{1}{\beta}\sum_{\sigma} \left[\log(1-\,e^{-\beta(\epsilon_{\sigma}- \mu)})+\log(1-\,e^{-\beta(\epsilon_{\sigma}+ \mu)})\right].
\end{eqnarray}
but also the alternative expression
\begin{equation}\label{integral}
  \mathcal{F}=\sum_{n=1}^{\infty}\left[(e^{n\beta\mu}+e^{-n\beta\mu})\right]c\int_{0}^{\infty}\frac{du}{\sqrt{4\pi}u^{3/2}}
  e^{-m^{2}c^{2}\left(\frac{(\beta m^{-1})^{2}n^{2}}{4u}+u\right)}\textrm{Tr}\,e^{-u(-\Delta_{\gamma}+U)}.
 \end{equation}
that will play an important role in the derivation of the nonrelativistic limit of $\mathcal{F}$. To the best of our knowledge in the literature on the ultrarelativistic expansion of the free energy in a static space-time \cite{Dowker3,Kirsten2} (\ref{freesum}) (or certain equivalent forms) is taken as the starting point. Here (and in the Appendix A) we gave a field theoretic derivation of (\ref{freesum}).

\subsection{Positivity Properties} \label{PosProp}

Our aim now is to examine the positivity of the operator $A$ given in (\ref{A0}). Since $c$ will play no particular role in this subsection we will set $c=1$.

Let us start by undoing our conformal transformation in
\begin{eqnarray}
  A &=& -\Delta_{\gamma}+\frac{d-1}{4d}R_{\gamma}+F\left(\xi-\frac{d-1}{4d}\right)R+F(m^{2}+V_{ext}).
\end{eqnarray}
In the operator language this corresponds to the similarity transformation
\begin{equation}
A\rightarrow A_{1}=F^{-\frac{d-1}{4}}AF^{\frac{d-1}{4}}.
\end{equation}
The result is
\begin{equation}\label{A1}
  A_{1}=\left[-\frac{F}{\sqrt{|g|}}\partial_{i}\sqrt{|g|}h^{ij}\partial_{j}+ F V\right].
\end{equation}
Let us now make one more similarity transformation
\begin{eqnarray}\label{A2}
  A_{2}&=& F^{-1/2}A_{1}F^{1/2}=F^{-\frac{d+1}{4}}AF^{\frac{d+1}{4}}\nonumber\\
  &=&F^{1/2}\left[-\frac{1}{\sqrt {|g|}}\partial_{i}\sqrt{|g|}h^{ij}\partial_{j}+V\right]F^{1/2}\nonumber\\
  &=&K_{2}+VF.
\end{eqnarray}
Here we defined
\begin{equation}\label{K2}
  K_{2}=F^{1/2}\left[-\frac{1}{\sqrt {|g|}}\partial_{i}\sqrt{|g|}h^{ij}\partial_{j}\right]F^{1/2}.
\end{equation}

In what follows we are going to assume $V\geq 0$. In particular if $V_{ext}\geq 0$  and the curvature coupling $\xi R$ vanishes, that is if we have either minimal coupling $\xi=0$ or a spacetime metric with $R=0$ we will have $V\geq 0$. For example any solution of the Einstein equation with vanishing cosmological constant and vanishing (or more generally  traceless) energy momentum tensor (for instance the Schwarzschild metric or its higher dimensional generalization Schwarzschild-Tangherlini metric) will automatically have $R=0$ and satisfy this condition.

If $A_{1}$ is defined on functions obeying the Neumann boundary condition then $A_{2}$ acts on functions subject to the Robin boundary condition
\begin{equation}\label{BC}
\left.N^{\mu} \partial_{\mu}(F^{1/2}f)\right|_{\partial B}=0\,\,\,\,\Leftrightarrow \,\,\,\, \left[N^{\mu} \partial_{\mu}f+(F^{-1/2}N^{\mu} \partial_{\mu}F^{1/2})f\right]_{\partial B}=0.
\end{equation}
Here $N=N^{\mu} \partial_{\mu}$ is the inward looking unit normal to $\partial B$. If on the other hand Dirichlet boundary condition is employed for $A_{1}$ then $A_{2}$ too is defined on functions obeying the Dirichlet boundary condition.

Consider the inner product
\begin{equation}
  (f_{1},f_{2})=\int_{B}d^{d}x\,\sqrt{|g|}f_{1}^{*}f_{2}.
\end{equation}
Now
\begin{eqnarray}\label{positivity}
  (f,A_{2}f)
  &=& \int d^{d}x \sqrt{|g|} f^{*} F^{1/2}\left[-\frac{1}{\sqrt{|g|}}\partial_{i}\sqrt{|g|}h^{ij}\partial_{j}+V\right]F^{1/2}f \nonumber \\
   &=& \int d^{d}x \sqrt{|g|}\left[h^{ij}\partial_{i} (fF^{1/2})^{*}\partial_{j} (fF^{1/2})+VFf^{*}f\right]\nonumber\\
   &&+\int_{\partial B} dS (fF^{1/2}) N_{i}h^{ij}\partial_{j}(fF^{1/2}).
\end{eqnarray}
Here the surface term vanishes because of the boundary conditions ((\ref{BC}) or Dirichlet) and the first term is obviously nonnegative. Thus we see that $ A_{2}$ and $A$ are positive operators. Finally one more partial integration shows that $A_{2}$ is a Hermitean operator.

Let us also note that (\ref{positivity}) also implies that $K_{2}$ is a positive operator. Moreover, in the case of Neumann/Robin boundary condition from (\ref{K2}) and (\ref{BC}) $F^{-1/2}$ is the ground state eigenfunction of $K_{2}$ and the lowest eigenvalue of $K_{2}$ is zero.

Now if we consider the operator inequalities
\begin{equation}
  K_{2}+(VF)_{min} \leq A_{2}=K_{2}+VF\leq  K_{2}+(VF)_{max}.
\end{equation}
and apply Rayleigh's variational principle we get
\begin{equation}
  \xi_{0}+(VF)_{min}\leq \epsilon_{0}^{2}\leq \xi_{0}+(VF)_{max}.
\end{equation}
Here $\xi_{0}$ is the lowest eigenvalue of $K_{2}$. As seen above $\xi_{0}=0$ for Neumann/Robin boundary conditions, and $\xi_{0}> 0$ for Dirichlet boundary conditions.

Also note that
\begin{eqnarray}
  (f_{1},A_{2}f_{2}) &=& (f_{1},F^{-\frac{d+1}{4}} AF^{\frac{d+1}{4}}f_{2}) \nonumber\\
   &=& \int_{B}d^{d}x \sqrt{|g|}\,F^{-\frac{d+1}{2}}\, (F^{\frac{d+1}{4}}f_{1})^{*} A(F^{\frac{d+1}{4}}f_{2})\nonumber\\
   &=& \int_{B}d^{d}x \sqrt{\gamma}\,\overline{\phi}_{1}^{*}A\overline{\phi}_{2}.
\end{eqnarray}
Thus the inner product induced by the similarity transformation is the natural one given by the Riemannian measure of the optical metric $\gamma_{ij}$
\begin{equation}\label{inner}
  \langle \overline{\phi}_{1},\overline{\phi}_{2}\rangle=\int_{B}d^{d}x \sqrt{\gamma}\,\overline{\phi}_{1}^{*}\overline{\phi}_{2}.
\end{equation}
Moreover, $A$ (together with Robin or Dirichlet boundary conditions) is a Hermitean operator relative to this inner product.

\section{Nonrelativistic Limit of the Free Energy}

In this section we will determine the large $c$ asymptotics of the free energy (\ref{integral}) by saddle point method. In the use of the saddle point method our treatment is similar in spirit to the treatment of the nonrelativistic Feynman propagator in a weak gravitational field given in \cite{Padmanabhan}. However in our analysis we will not make any weak field assumption in the sense of \cite{Padmanabhan}.

In what follows, we will keep all the factors of $c$ explicit in the calculations. It will be convenient to use the path integral representation of the trace term in (\ref{integral})
\begin{eqnarray}\label{pathtr}
  Tr\,e^{-u(-\Delta_{\gamma}+U)} &=& \int\mathcal{D}x e^{-s(u,c)}
\end{eqnarray}
Here
\begin{equation}
  s(u,c)=\int_{0}^{u}d\tau\,\left[\frac{1}{4}\gamma_{ij}\frac{dx^{i}}{d\tau}\frac{dx^{j}}{d\tau}+U(x(\tau))\right],
\end{equation}
and the path integral is taken over the closed paths.

So using this representation of trace in (\ref{integral}) we get
\begin{equation}\label{integral2}
  \mathcal{F}=\sum_{n=1}^{\infty}\left[2\cosh(n\beta\mu)\right]  \int\mathcal{D}x\int_{0}^{\infty}\frac{cdu}{u^{3/2}}\frac{1}{\sqrt{4\pi}}
  e^{-m^{2}c^{2}\left(\frac{(\beta m^{-1})^{2}n^{2}}{4u}+u\right)}e^{-s(u,c)}.
 \end{equation}
 The $u$ integral in this expression is therefore of the form
\begin{equation}\label{integral1}
  \int_{0}^{\infty}\,\frac{cdu}{\sqrt{4\pi}u^{3/2}} \,e^{-s(u,c)}e^{-c^{2}r(u)},
\end{equation}
where
\begin{equation}\label{rn}
  r(u)=m^{2}\left( \frac{n^{2}(\beta m^{-1})^{2}}{4u}+u\right).
\end{equation}
Our strategy is to apply Laplace method to evaluate the large $c$ limit of this integral.

\subsection{Nonrelativistic Limit in the Ultrastatic Case }
Before we examine the nonrelativistic limit of $\mathcal{F}$ in a static spacetime let us examine the simpler case of ultrastatic spacetime $\mathbf{M^{(d+1)}}=\mathbf{R}\times
\mathbf{M^{(d)}}$, where $\mathbf{R}$ is the global time and $\mathbf{M^{(d)}}$ is a Riemannian manifold with metric $h_{ij}$.  The metric on $\mathbf{M^{(d)}}$ is
\begin{equation}
  ds^{2}=-c^{2}dt^{2}+h_{ij}(x)dx^{i}dx^{j}.
\end{equation}
So, $F=1$ and $\gamma_{ij}=h_{ij}$. Here typically $h_{ij}$ is independent of $c$ and so is the trace term $s(u,c)=s(u)$.

Then in the large $c$ limit we have
\begin{equation}\label{mainust}
 \int \frac{cdu}{\sqrt{4\pi}u^{3/2}}\, e^{-s(u)}e^{-c^{2}r(u)}\sim \frac{c}{\sqrt{4\pi}\overline{u}^{3/2}}\,e^{-s(\overline{u})}e^{-c^{2}r(\overline{u})}\sqrt{\frac{2\pi}{c^{2}r''(\overline{u})}}.
\end{equation}
Now the saddle point $\overline{u}$ is given by
\begin{equation}\label{saddle}
  \frac{d}{du}\left[\frac{(n\beta m^{-1})^{2}}{4u}+u\right]=0,
\end{equation}
as
\begin{equation}
  \overline{u}=\frac{n\beta}{2m}.
\end{equation}
So
\begin{equation}\label{par}
  c^{2}r(\overline{u})=n\beta mc^{2},\;\;\;\;\;\;\;
 c^{2}r''(\overline{u})=\frac{(cn\beta)^{2}}{2\overline{u}^{3}}.
\end{equation}
Using (\ref{mainust}) and (\ref{par})  in the integral in (\ref{integral2}) we have
\begin{equation}
 \int\mathcal{D}x \int \frac{cdu}{\sqrt{4\pi}u^{3/2}}\, e^{-s(u)}e^{-c^{2}r(u)}\sim \frac{e^{-n\beta mc^{2}}}{n\beta}\int\mathcal{D}x  e^{-s(\overline{u})}
\end{equation}
Combining with (\ref{pathtr}) we get
\begin{equation}
 \frac{e^{-n\beta mc^{2}}}{n\beta}\int\mathcal{D}x  e^{-s(\overline{u})} =
 \frac{e^{-n\beta mc^{2}}}{n\beta}Tr\,e^{-n\beta\, \left[\frac{1}{2m}(-\Delta_{h}+U)+mc^{2}\right]}
\end{equation}

We now assume that the background is not strong enough to cause pair production \cite{zeldovich}. We will be more quantitative about this assumption in Sec. 3.2 when we discuss the static background which is the case of physical interest. Since the number of particles and antiparticles are separately conserved, one can study particle and antiparticle thermodynamics separately. Focusing on the particle thermodynamics (treatment of antiparticles being similar) we get
\begin{eqnarray}
 \mathcal{F}_{NR} &=&-\sum_{n=1}^{\infty}\frac{1}{n\beta}\,Tr \,e^{-n \beta (L-\mu)}\\
  &=& \frac{1}{\beta}\sum_{\sigma}\log(1-e^{-\beta( \lambda_{\sigma}-\mu)}).
\end{eqnarray}
Here
\begin{equation}\label{LU0}
  L=\frac{1}{2m}(-\Delta_{h}+U)+mc^{2},
\end{equation}
and $\left\{\lambda_{\sigma}\right\}$ is the spectrum of $L$. The chemical potential must satisfy $\mu<\lambda_{0}$, where $\lambda_{0}$ is the lowest eigenvalue of $L$. The operator $L$ is in fact the Schr\"{o}dinger operator $H_{NR}$ corresponding to the nonrelativistic limit of the KG equation. So we recover the expected result for the free
 energy of a nonrelativistic system on the space manifold $\mathbf{M^{(d)}}$. However in Sec. 3.3 we will see that in a static background the relation between $L$ and $H_{NR}$, and in fact the determination of $H_{NR}$, are not immediate and require further analysis.

\subsection{Nonrelativistic Limit in the Static Case}

Here we will work explicitly in the Schwarzschild spacetime for which
\begin{equation}
 F=1-\frac{2GM}{c^{2}r}
\end{equation}
and
\begin{equation}\label{Sch}
    ds^{2}=-\left(1-\frac{r_{s}}{r}\right)c^{2}dt^{2}+\left(1-\frac{r_{s}}{r}\right)^{-1}dr^{2}+r^{2}(d\theta^{2}+\sin^{2}\theta d\phi^{2}).
\end{equation}
Here $r_{s}$ is the Schwarzschild radius
\begin{equation}
  r_{s}=\frac{2MG}{c^{2}}.
\end{equation}
 In dealing with the thermodynamics of a quantum field near a black hole one encounters divergences in thermodynamic quantities as one approaches the horizon \cite{Barbon}. The existence and properties of these horizon divergences are well known in the ultrarelativistic limit. We expect them to persist in the nonrelativistic limit as well. In what follows we will stay away from the horizon by confining the quantum field in a region away from the horizon.  We plan to come back to the question of horizon divergences in a future work. For the sake of definiteness one may consider a spherical shell $B$ of inner radius $r_{1}$ and outer radius $r_{2}$. The outer radius will provide an infrared cutoff for the system while the inner radius will keep the system away from the horizon. In a singularity free geometry the inner wall may represent the surface of the gravitating object.

Before we apply the saddle point method to (\ref{integral}) in a static spacetime we must examine the factors of $c$ in $-\Delta_{\gamma}+U$.

The optical metric corresponding to (\ref{Sch}) is
\begin{equation}
  \gamma_{ij}=diag\left(\frac{1}{\left(1-\frac{r_{s}}{r}\right)^{2}},\frac{r^{2}}{\left(1-\frac{r_{s}}{r}\right)},
  \frac{r^{2}\sin^{2}\theta}{\left(1-\frac{r_{s}}{r}\right)}\right).
\end{equation}
Observe that for large $c$
\begin{equation}
  \gamma_{ij}=diag\left(1+O(c^{-2}),r^2[1+O(c^{-2})],
  r^2\sin^{2}\theta[1+O(c^{-2})]\right),
\end{equation}
or passing to Cartesian coordinates
\begin{equation}
  \gamma'_{ij}= \delta_{ij}+O(c^{-2}).
\end{equation}

The scalar curvature of $\gamma_{ij}$ is
\begin{equation}
  R_{\gamma}=-\frac{6M^{2}}{c^{4}r^{4}}=O(c^{-4}),
\end{equation}
and from $F=1+O(c^{-2})$ we also get
\begin{equation}
  (F-1)(mc)^{2}=2m^{2}\Phi=O(c^{0}),
\end{equation}
Here $\Phi=-M/r$ is the Newtonian gravitational potential.

These observations together with the fact that $R=0$ for the Schwarzschild metric imply
\begin{equation}
  U=O(c^{-4}).
\end{equation}
So,
\begin{equation}\label{suc}
s(u,c)=\int d\tau\,\left[\frac{1}{4}\delta_{ij}\frac{dx^{i}}{d\tau}\frac{dx^{j}}{d\tau}+2m^{2}\Phi\right]+O(c^{-2})=s(u,c=\infty)+O(c^{-2}).
\end{equation}
Since $s(u,c)$ is of smaller order in $c$ than $c^{2}r(u)$ the generalized Laplace method \cite{Copson} asserts that the integral is localized around the minimum of the latter and we get
\begin{equation}\label{mainst}
 \int \frac{cdu}{\sqrt{4\pi}u^{3/2}}\, e^{-s(u,c)}e^{-c^{2}r(u)}\sim \frac{c}{\sqrt{4\pi}\overline{u}^{3/2}}\,e^{-s(\overline{u},c)}e^{-c^{2}r(\overline{u})}\sqrt{\frac{2\pi}{c^{2}r''(\overline{u})}}.
\end{equation}

If desired this expression may be further simplified by replacing $s(\overline{u},c)$ by $ s(\overline{u},c=\infty)$ which just leads to the free energy of an ideal gas in the classical Newtonian potential $m\Phi$. However the main point here is that one does not have to do that simplification; as it stands (\ref{mainst}) is an improved asymptotics over the simplified version and is the one we shall use. Most importantly (\ref{mainst}) captures the background geometry; replacing $s(\overline{u},c)$ by $ s(\overline{u},c=\infty)$ means losing the background geometry and this is precisely what we want to avoid. In this sense our large $c$ limit may be regarded as a post-Newtonian approximation to the free energy where the background geometry is treated to all orders in $c^{-1}$.

From (\ref{suc}) we see that the coupling energy between the background and the matter is $m\Phi+O(c^{-2})$. The pair production cannot occur if $|m\Phi(r_{1})|+O(c^{-2}) < 2mc^{2}$ which is certainly the case in the large $c$ nonrelativistic regime, unless $|m\Phi(r_{1})|$ is very large. So assuming we are not too close to the gravitation center and the gravitating object is not supermassive we can ignore pair production and just focus on the particle thermodynamics (treatment of the antiparticles being similar).

Thus again defining
\begin{eqnarray} \label{L}
 L &=& \frac{1}{2m}\left(-\Delta_{\gamma}+U\right)+mc^{2}.
\end{eqnarray}
with eigenvalues $\{\lambda_{\sigma}\}$, we get
\begin{eqnarray}\label{fnrel}
  \mathcal{F}_{NR}&=&-\sum_{n=1}^{\infty}\frac{1}{n\beta}\,Tr\,e^{-n\beta(L-\mu)}=\frac{1}{\beta} Tr \log\left[1-e^{-\beta(L-\mu)}\right].
\end{eqnarray}
Moreover, the particle number is given by the Bose-Einstein distribution
\begin{eqnarray}\label{Nnrel}
N=-\frac{\partial \mathcal{F_{NR}}}{\partial \mu} = \sum_{\sigma} \frac{1}{e^{\beta(\lambda_{\sigma}-\mu)}-1}.
\end{eqnarray}

So the nonrelativistic limit of the thermodynamics on a static spacetime manifold is governed by the operator $L$ defined on the optical manifold. In the next section we will discuss the relation of $L$ to the nonrelativistic limit of the Klein-Gordon equation. Finally let us note that although we worked with the specific example of Schwarzschild metric our discussion in this section is easily seen to be valid for static metrics for which $(F-1)(mc)^{2}=O(c^{0})$ for large $c$.

\subsection{$L$ vs. $H_{NR}$}

Consider the KG equation
\begin{equation}\label{KGEQN}
 ( -\Delta_{g}+\xi R+(mc)^{2}+V_{ext})\phi=0
\end{equation}
In an ultrastatic spacetime ($F=1$) this can be written explicitly as
\begin{equation}
  -\partial_{0}^{2}\phi= (-\Delta_{h}+\xi R+V_{ext}+(mc)^{2})\phi.
\end{equation}
The nonrelativistic limit can be obtained by taking the square root of the operator appearing
on the right hand side and then expanding the result in inverse powers of $(mc)^{2}$. The result is the Sch\"{o}dinger equation
\begin{equation}
  i\partial_{0}\phi=H_{NR}\phi,
\end{equation}
where
\begin{equation}
  H_{NR}=\frac{1}{2m}(-\Delta_{h}+U)+mc^{2},
\end{equation}
and this obviously coincides with $L$ given in (\ref{LU0}).

On the other hand in a general static spacetime the KG equation is written explicitly as
\begin{equation}\label{Klein1}
  -\partial_{0}^{2}\phi=-\left(\frac{F}{\sqrt{|g|}}\partial_{i}\sqrt{|g|}h^{ij}\partial_{j}+F(mc)^{2}+\xi FR+FV_{ext}\right)\phi=0.
\end{equation}
On a general static spacetime the above argument for the ultrastatic case does not work because the mass term $F(mc)^{2}$ does not commute with the remaining operators on the right hand side of (\ref{Klein1}). In general, finding the nonrelativistic limit of the KG equation on a curved spacetime is a difficult task (see \textit{e.g.} \cite{deWitt, Cognola}). However our saddle point argument translates into a simple, albeit formal, nonrelativisitic limit of the KG equation in the type of static spacetimes we consider in this work. Recalling the
counting argument of Sec. 2.2 we write the mass term as $(F-1)(mc^{2})+(mc)^{2}$ and assume $(F-1)(mc)^{2}=O(c^{0})$. Then the right hand side of (\ref{Klein1}) can be written as
\begin{equation}
  c^{2}\left[-\frac{F}{\sqrt{|g|}}\partial_{i}\sqrt{|g|}h^{ij}\partial_{j}+(F-1)(mc)^{2}+\xi R F+(mc)^{2}\right].
\end{equation}
where all the terms in square brackets except $(mc)^{2}$ involve negative powers of $c$ and are therefore $O(c^{0})$.

Now we can proceed as in the ultrastatic case and expand the formal square root of this operator in inverse powers of $c^{2}$ or equivalently of $(mc)^{2}$. Thus we get
\begin{eqnarray}
  H_{NR}&=&mc^{2}+\frac{1}{2m}\left[-\frac{F}{\sqrt{|g|}}\partial_{i}\sqrt{|g|}h^{ij}\partial_{j}+U\right]+\ldots\\
  &=& mc^{2}+\frac{1}{2m}\left(c^{-2}A_{1}-(mc)^{2}\right)+O((c^{-2}A_{1})^{2})
\end{eqnarray}
with $A_{1}$ given as in (\ref{A1}). As we saw in Sec. 3 $A_{1}$ is related to $A$ by the similarity transformation $A=F^{\frac{d-1}{4}}A_{1}F^{-\frac{d-1}{4}}$, so the above operator
is similar to
\begin{eqnarray}
 F^{\frac{d-1}{4}}H_{NR}F^{-\frac{d-1}{4}}= mc^{2}+\frac{1}{2m}\left(c^{-2}A-(mc)^{2}\right)= mc^{2}+\frac{1}{2m}\left(-\Delta_{\gamma}+U\right)=L.
\end{eqnarray}
Thus we conclude that $L$ and $H_{NR}$ are related by a similarity transformation.

\section{Expansions of the Free Energy and the Occupation Number}

In this section we will derive asymptotic expansions for the free energy and the occupation number which will be used in the analysis of condensation. The expansion parameter will be presented shortly.

Let us start by writing  (\ref{fnrel}) as
\begin{eqnarray}\label{harm1}
  \mathcal{F}_{NR}&=&-\sum_{n=1}^{\infty}\frac{1}{n\beta}\,Tr\,e^{-n\beta(\lambda_{0}-\mu)}e^{-n\beta(\lambda_{0}-\mu)\widetilde{L}}.
\end{eqnarray}
where
\begin{equation}
  \widetilde{L}=\frac{(L-\lambda_{0})}{\lambda_{0}-\mu}.
\end{equation}
This expression is in the form of a harmonic sum
\begin{equation}\label{harm}
  \mathcal{F}(x)=\sum_{n=1}^{\infty}h(nx),
\end{equation}
where $x=\beta(\lambda_{0}-\mu)$ and is given in terms of the heat kernel of $\widetilde{L}$ as
\begin{equation}\label{harm2}
  h(x)=-(\lambda_{0}-\mu)\,\frac{e^{-x}}{x}\;Tr\,e^{-x\widetilde{L}}.
\end{equation}
By using Mellin transform techniques, we will analyze this harmonic sum in the small $x$ limit, which is the relevant regime for the Bose-Einstein condensation \cite{Ober,Wng,Flajolet2}. In what follows we will drop the subscript $NR$ from $\mathcal{F}_{NR}$.

The Mellin transform of a function $g(x)$ is defined as
\begin{equation}
  \widetilde{g}(s)=\int_{0}^{\infty}dx\,x^{s-1}g(x),\;\;\;s\in \mathbf{C}.
\end{equation}
If we have a meromorphic extension of $ \widetilde{g}(s)$ with the singular part
\begin{equation}
\widetilde{g}(s)\asymp\sum_{w,k}\frac{Res(w,k)}{(s-w)^{k+1}},
\end{equation}
where $\asymp$ refers to the singular part of $\widetilde{g}(s)$, then the asymptotic behavior of the function itself is given by \cite{Ober,Wng,Flajolet2}
\begin{equation}\label{asymp}
 g(x)\sim \sum_{w,k}Res(w,k)\frac{(-1)^{k}}{k!}x^{-w}(\log x)^{k}.
\end{equation}
The validity of (\ref{asymp}) for the more difficult case where $h$ involves the Poisson kernel instead of the heat kernel was shown in the Appendix B of \cite{BECUS}. Our easier case of heat kernel dependent $h$ can be treated by straightforwardly adapting the discussion given there for the Poisson kernel. We will omit the details of that treatment here and use (\ref{asymp}) without reserve in what follows.

Taking the Mellin transform of $\mathcal{F}(x)$ we get
\begin{equation}
  \widetilde{\mathcal{F}}(s)=\zeta(s)\widetilde{h}(s).
\end{equation}
where
\begin{equation}
\zeta(s)= \sum_{k=1}^{\infty} k^{-s}
\end{equation}
is the Riemann-Zeta function and
\begin{equation}\label{hs}
  \widetilde{h}(s)=-(\lambda_{0}-\mu)\int_{0}^{\infty}dx\,x^{s-2}\,e^{-x }Tr\,e^{-x\widetilde{L}}.
\end{equation}
By the decay properties of the heat kernel the $x$ integral is well behaved in the upper limit of integration. The divergence comes from the vicinity of $x=0$ and the meromorphic continuation is obtained by subtracting and adding the terms of the heat kernel expansion of $\widetilde{L}$,
\begin{equation} \label{HKE}
 Tr  e^{-x\, \widetilde{L} }\sim\sum_{j=0}^{\infty} \widetilde{a}_{j/2}x^{\frac{j-d}{2}}.
\end{equation}
For example since as $x\rightarrow 0$, $Tr\,e^{-x\widetilde{L}}\sim \widetilde{a}_{0}x^{-d/2}$, initially the $x$ integral is convergent for $\textrm{Re}s>1+d/2$ but if we subtract and add the leading term of the heat kernel expansion we get
\begin{equation}
  \int_{0}^{\infty}dx\,x^{s-2}\,e^{-x }\left[Tr\,e^{-x\widetilde{L}}-\frac{\widetilde{a}_{0}}{x^{d/2}}\right]+\Gamma\left(s-1-\frac{d}{2}\right).
\end{equation}
Since the difference between the trace and the leading term of its heat kernel expansion is $O(x^{-(d-1)/2})$ the integral is now convergent and holomorphic for $\textrm{Re}s>1+(d-1)/2$, and the gamma function is meromorphic. Continuing in this manner we extend $\widetilde{h}(s)$ to successively larger regions to the left of the original convergence region $\textrm{Re}s>1+d/2$.

Using (\ref{HKE}) in (\ref{hs}) we get the divergent piece of $ \widetilde{h}(s)$ as
\begin{equation}
-(\lambda_{0}-\mu)\sum_{j=0}^{\infty} \widetilde{a}_{j/2}\Gamma\left(s-1+\frac{j-d}{2}\right).
\end{equation}
Since $\Gamma(x)$ has the singular expansion
\begin{equation}
  \Gamma(x)\asymp \sum_{l=0}^{\infty} \frac{(-1)^{l}}{l!}\frac{1}{x+l},
\end{equation}
we have
\begin{equation} \label{htild}
   \widetilde{h}(s)\asymp -(\lambda_{0}-\mu)\sum_{j,l} \widetilde{a}_{j/2}\frac{(-1)^{l}}{l!}\frac{1}{s-1+\frac{j-d}{2}+l}.
\end{equation}
On the other hand $\zeta(s)$ has a unique simple pole at $s=1$ and
\begin{equation}\label{zeta}
\zeta(s)\sim \frac{1}{s-1}+\gamma,\;\;\;\;\;\;s\rightarrow 1
\end{equation}
So all the poles of $\widetilde{\mathcal{F}}(s)=\zeta(s)\widetilde{h}(s)$ are simple except the double pole at $s=1$. The set of all poles is $\mathcal{I}=\{(d+2-n)/2:n=0,1,2,\ldots\}$. The corresponding residues are given as $ -(\lambda_{0}-\mu)c_{n/2}$, where
\begin{equation}\label{cn2}
  c_{n/2}=\sum_{l=0}^{[|\frac{n}{2}|]}\frac{(-1)^{l}}{l!}\widetilde{a}_{\frac{n}{2}-l}.
\end{equation}
The coefficients $c_{n/2}$ are readily seen to be the heat kernel coefficients of  $\widetilde{L}+1=(L-\mu)/(\lambda_{0}-\mu)$
\begin{equation}
  \textrm{Tr} e^{-x(\widetilde{L}+1)}\sim \sum_{n=0}^{\infty}c_{n/2}x^{\frac{-d+j}{2}}.
\end{equation}

We can now use (\ref{asymp}) and obtain the asymptotic expansion of $\mathcal{F}$. For convenience, we separate the expansion into two parts as
\begin{equation}
\mathcal{F}(x)\sim \mathcal{F}_s(x)+\mathcal{F}_d(x),
\end{equation}
where $\mathcal{F}_s(x)$ denotes terms coming from the simple poles when $s\neq1$ and  $\mathcal{F}_d(x)$ is the contribution of the double pole at $s=1$.

Let us first consider $\mathcal{F}_s$ which is given as
\begin{equation}\label{FSW}
  \mathcal{F}_{s}(x)= -(\lambda_{0}-\mu)\sideset{}{'}\sum_{n=0}^{\infty}\zeta\left(\frac{d+2-n}{2}\right)c_{n/2}x^{-\frac{d+2-n}{2}}.
\end{equation}
Here prime on the summation sign means $n=d$ term corresponding to the double pole at $s=1$ is omitted in the sum.

The contribution coming from $s=1$ ($n=d/2$), which is the double pole of $\widetilde{\mathcal{F}}(s)$, is calculated from the following $s\rightarrow 1$ asymptotic behaviour
\begin{align} \nonumber
 \widetilde{\mathcal{F}}(s)& \sim \left (\frac{1}{s-1}+\gamma \right )
\left(\frac{res(1,0)}{s-1}  +R_{+}\tilde{h}(s=1) \right) \\
 &\sim\frac{res(1,0)}{(s-1)^2} + \frac{\gamma res(1,0)+R_{+}\tilde{h}(s=1)}{s-1}.
\end{align}
Here $res(1,0)=-(\lambda_{0}-\mu)c_{d/2}$ is the residue of $\tilde{h}(s)$ at $s=1$ and $R_{+}\tilde{h}(s)$ means the holomorphic part of $\tilde{h}(s)$.

Using (\ref{asymp}) we obtain
\begin{eqnarray}
 \mathcal{F}_{d}(x) &=& -(\lambda_{0}-\mu)\left\{\left[c_{d/2}\gamma-\frac{R_{+}(\tilde{h}(1))}{\lambda_{0}-\mu}\right]x^{-1}-c_{d/2}x^{-1}\log x\right\}
\end{eqnarray}
Let us now note that
\begin{equation}
  \widetilde{h}(s)=-(\lambda_{0}-\mu)\Gamma(s-1)\zeta_{\widetilde{L}+1}(s-1).
\end{equation}
Since
\begin{equation}
  \Gamma(s-1)\sim \frac{1}{s-1}-\gamma\;\;\;\;\;\;s\rightarrow 1,
\end{equation}
as $s\rightarrow 1$ we have
\begin{eqnarray}
  \zeta(s) \widetilde{h}(s)&\sim &-(\lambda_{0}-\mu)  \left(\frac{1}{s-1}+\gamma \right ) \left(\frac{1}{s-1}-\gamma \right ) \left(\zeta_{\widetilde{L}+1}(0)+\zeta'_{\widetilde{L}+1}(0)(s-1) \right )\nonumber\\
  &\sim &-(\lambda_{0}-\mu)\left[\frac{\zeta_{\widetilde{L}+1}(0)}{(s-1)^{2}}+\frac{\zeta'_{\widetilde{L}+1}(0)}{s-1}\right].
\end{eqnarray}
So
\begin{equation}
  \mathcal{F}_{d}(x)=-(\lambda_{0}-\mu)\left[ \zeta_{\widetilde{L}+1}(0) \frac{T}{\lambda_{0}-\mu}\log\left(\frac{T}{\lambda_{0}-\mu}\right)+\zeta'_{\widetilde{L}+1}(0)\frac{T}{\lambda_{0}-\mu}\right].
\end{equation}
In this form our expansion $\mathcal{F}\sim \mathcal{F}_s+\mathcal{F}_d$, including the spectral zeta function term, is the direct nonrelativistic analog of the ultrarelativistic (high temperature) expansion derived by Dowker and Kennedy in \cite{Dowker1}.

Let us now turn to the particle number which by (\ref{harm1}) is given as
\begin{eqnarray}\label{nharm1}
  N=-\frac{\partial\mathcal{F}}{\partial\mu}=\sum_{p=1}^{\infty} Tr\,e^{-p\beta(\lambda_{0}-\mu)}e^{-p\beta(\lambda_{0}-\mu)\widetilde{L}}.
\end{eqnarray}
So again we have a harmonic sum
\begin{equation}\label{nharm}
  N(x)=\sum_{p=1}^{\infty}n(px),
\end{equation}
with
\begin{equation}\label{nharm2}
  n(x)= e^{-x}\;Tr\,e^{-x\widetilde{L}}.
\end{equation}
Taking the Mellin transform of $N(x)$ we get
\begin{equation}
  \widetilde{N}(s)=\zeta(s)\widetilde{n}(s).
\end{equation}
and
\begin{equation}\label{ns}
  \widetilde{n}(s)=\int_{0}^{\infty}dx\,x^{s-1}\,e^{-x }Tr\,e^{-x\widetilde{L}},
\end{equation}
Using the heat kernel expansion we get the divergent part of $\widetilde{n}(s)$
\begin{equation}
 \sum_{j=0}^{\infty} \widetilde{a}_{j/2}\Gamma\left(s+\frac{j-d}{2}\right).
\end{equation}
So the singular expansion is
\begin{equation} \label{ntild}
   \widetilde{n}(s)\asymp \sum_{j,l} \widetilde{a}_{j/2}\frac{(-1)^{l}}{l!}\frac{1}{s+\frac{j-d}{2}+l}.
\end{equation}
Now the poles are given as $\mathcal{J}=\{(d-n)/2:n=0,1,2,\ldots\}$ and the corresponding residues as $c_{n/2}$. Again the pole at $s=1$ is double and we separate $N$ as
$N(x)=N_{s}(x)+N_{d}(x)$.

A repetition of the analysis for the free energy yields for $N$
\begin{equation}\label{nsingle}
  N_{s}(x)=\sideset{}{'}\sum_{n=0}^{\infty} \zeta\left(\frac{d-n}{2}\right) c_{n/2}x^{\frac{n-d}{2}},
\end{equation}
and
\begin{eqnarray}\label{ndouble}
  N_{d}(x)=[\gamma c_{(d-2)/2}+R_{+}\widetilde{n}(s=1)]x^{-1}-c_{(d-2)/2}x^{-1}\log x.
\end{eqnarray}
Here the prime on the summation means that $n=d-2$ term corresponding to the double pole is omitted in the summation for $N_{s}$.

Now $c_{n/2}$'s are the heat kernel coefficients of
\begin{equation}
 \widetilde{L}+1=\frac{L-\mu}{\lambda_{0}-\mu}=\frac{-\Delta_{\gamma}+U+2m^{2}c^{2}-2m\mu}{2m(\lambda_{0}-\mu)}.
\end{equation}
It will be useful to express $c_{n/2}$'s in terms of the heat kernel coefficients $a_{n/2}$ of $-\Delta_{\gamma}+U+2m^{2}c^{2}-2m\mu$ which involve geometric invariants of the optical metric. This can be done easily by employing the scaling properties of heat kernel coefficients \cite{Gilkeybook} which leads to
\begin{equation}\label{cvea}
  c_{n/2}=\frac{a_{n/2}}{(2m(\lambda_{0}-\mu))^{\frac{n-d}{2}}}.
\end{equation}

\section{Bose Einstein Condensation}

 We will now examine the gravitational effects on Bose-Einstein condensation in a region $B$ with a large but finite volume in a static background away from any horizon. We will focus on $d=3$ and work explicitly in the Schwarzschild background. Thermodynamic densities will be calculated by dividing extensive quantities by the proper volume $V_{prop}$ of $B$ which is the volume measured by a static observer (fiducial observer \cite{Thorne}),
 \begin{equation}\label{vprop}
   dV_{prop}=\frac{r^{2}}{\sqrt{F}}\sin\theta dr d\theta d\phi.
 \end{equation}
 In what follows when we calculate the heat kernel coefficients explicitly we will assume that the region $B$ is the spherical shell of inner radius $r_{1}$ and outer radius $r_{2}$. Then, as we will see shortly, $V_{prop}=O(r_{2}^{3})$ for $r_{2}\rightarrow \infty$.

 Since
 \begin{equation}
   L=-\frac{\delta^{\mu\nu}\partial_{\mu}\partial_{\nu}}{2m}+O(c^{-2}),
 \end{equation}
 the gap between the eigenvalues of $L$ is expected to be $O(r_{2}^{-2})$ (plus perturbative corrections in inverse powers of $c$). If $\lambda_{0}-\mu=O(r_{2}^{-3})$ then from
 \begin{equation}\label{no}
 N_{0}=\frac{1}{e^{\beta(\lambda_{0})-\mu}-1}\simeq \frac{T}{\lambda_{0}-\mu}
 \end{equation}
 we see that in the thermodynamic limit $N_{0}/V_{prop}\neq 0$. Therefore $\lambda_{\sigma}-\mu=(\lambda_{0}-\mu)+(\lambda_{\sigma}-\lambda_{0})$ is $O(r_{2}^{-2})$. This implies $N_{\sigma}/V_{prop}\rightarrow 0$ and condensation to the ground state occurs in the thermodynamic limit. In a finite volume condensation does not occur. This is because just above the critical temperature one cannot set $N_{0}=0$ since that would yield infinite $\lambda_{0}-\mu$. Below we will investigate the dependence of the temperature on the depletion coefficient, \textit{i.e.} the number density of excited particles, in a large but finite volume and derive the gravitational/geometric effects on this relation.
 To do this we must isolate the excited state contribution $N_{e}$ to $N$. This can be done by omitting the ground state contribution in the trace term in (\ref{nharm1}). We shall denote the resulting trace by $Tr'$. Thus our expansion for $N$ can be used for $N_{e}$ after replacing in it all the heat kernel coefficients of $\mathrm{Tr}e^{-x\widetilde{L}}$ by those of $\mathrm{Tr}'e^{-x\widetilde{L}}$.

\subsection{Leading Order}

Consider (\ref{nsingle}) to the leading order.
\begin{eqnarray}\label{N}
  N_{e}=a_{0}\zeta\left(\frac{3}{2}\right)(2mT)^{3/2}=\left(\frac{m}{2\pi}\right)^{3/2}\zeta\left(\frac{3}{2}\right)V_{\gamma}T^{3/2}.
\end{eqnarray}
Here we used (\ref{cvea}) to write $c_{0}$ in terms of $a_{0}$ which is given as \cite{Gilkey,Vass}
\begin{equation}
   a_{0} = \frac{1}{(4\pi)^{3/2}}\int_{B} dV_{\gamma}
\end{equation}
with the volume form of the optical metric given as
\begin{equation}
  dV_{\gamma}=(rF^{-1})^{2}\sin^{2}\theta dr d\theta d\phi.
\end{equation}
Solving (\ref{N}) for $T$ we get
\begin{eqnarray}\label{T0}
  T&=&\frac{2\pi}{m}\left(\frac{N_{e}}{\zeta\left(\frac{3}{2}\right)V_{\gamma}}\right)^{2/3}\\
  &=&\frac{2\pi}{m}\left(\frac{n_{e}}{\zeta\left(\frac{3}{2}\right)}\right)^{2/3}\left(\frac{V_{prop}}{V_{\gamma}}\right)^{2/3}.
\end{eqnarray}
Now for the Schwarzschild metric we have
\begin{equation}
  F= 1-\frac{r_{s}}{r}.
\end{equation}
All integrals are elementary but since we are interested in a large but finite volume we shall evaluate them asymptotically.
\begin{eqnarray}\label{vol}
  V_{\gamma}&=&\int d^{3}x\,\sqrt{\gamma}=4\pi\int_{r_{1}}^{r_{2}}dr\,\frac{r^{2}}{F^{2}}\sim  4\pi\int_{r_{1}}^{r_{2}}dr\,r^{2}\left(1+\frac{2r_{s}}{r}+\ldots\right)\nonumber\\
   &\sim & 4\pi\left(\frac{r^{3}_{2}}{3}+r_{s}r_{2}^{2}+\ldots\right).
\end{eqnarray}
Similarly,
\begin{eqnarray}\label{vrop}
  V_{prop}=4\pi\int_{r_{1}}^{r_{2}}dr\,\frac{r^{2}}{\sqrt{F}}\sim  4\pi \left(\frac{r^{3}_{2}}{3}+\frac{r_{s}r_{2}^{2}}{4}+\ldots\right).
\end{eqnarray}
So
\begin{equation}\label{vratio}
  \frac{V_{prop}}{V_{\gamma}}\sim 1-\frac{9r_{s}}{4r_{2}}.
\end{equation}
Thus
\begin{eqnarray}\label{T0ac}
  T=\frac{2\pi}{m}\left(\frac{n_{e}}{\zeta\left(\frac{3}{2}\right)}\right)^{2/3}\left(1-\frac{3r_{s}}{2r_{2}}\right).
\end{eqnarray}
For $r_{s}=0$ this reduces to the flat space result \cite{Pathria}. Also in the thermodynamic limit $r_{2}\rightarrow\infty$ we get the flat space result. Thus the geometric/gravitational effects are washed out in the thermodynamic limit. Heuristically this is not so surprising since the spacetime is asymptotically flat and in the thermodynamic limit the largest contribution to the (convergent) integral comes from the flat region.

Let us now examine the equation of state using (\ref{FSW}) and (\ref{cvea}). At the leading order we have
\begin{eqnarray}
 P &=& -\frac{\partial \mathcal{F}}{\partial V_{prop}}=\frac{1}{2m}\zeta\left(\frac{5}{2}\right)\frac{\partial a_{0}}{\partial V_{prop}}(2mT)^{5/2}\nonumber\\
 &=&\frac{1}{2m}\zeta\left(\frac{5}{2}\right)\frac{1}{(4\pi)^{3/2}}\frac{dV_{\gamma}}{d V_{prop}}(2mT)^{5/2}.
\end{eqnarray}
So,
\begin{eqnarray}
  \frac{P}{n}=\frac{\zeta\left(\frac{5}{2}\right)}{\zeta\left(\frac{3}{2}\right)}\frac{V_{prop}}{V_{\gamma}}\frac{dV_{\gamma}}{d V_{prop}}T
  \end{eqnarray}
But
\begin{equation}
  \frac{dV_{\gamma}}{dV_{prop}}=\frac{dV_{\gamma}/dr_{2}}{dV_{prop}/dr_{2}}=F^{-3/2}(r_{2})\sim 1+\frac{3r_{s}}{2r_{2}}.
\end{equation}
 So we get
  \begin{eqnarray}
  \frac{P}{n} =\frac{\zeta\left(\frac{5}{2}\right)}{\zeta\left(\frac{3}{2}\right)}\left(1-\frac{3}{4}\frac{r_{s}}{r_{2}}\right)T.
\end{eqnarray}
Again for $r_{s}=0$ or $r_{2}\rightarrow \infty$ this reduces to the flat space equation of state near the critical temperature \cite{Pathria}.

\subsection{Boundary Effects}

Let us now consider the effects of the sub-leading term in the expansion of $N_{e}$. At $d=3$ this term is the logarithmic term in $N_{d}$ (\ref{ndouble}).
\begin{equation}\label{ne2}
  N_{e}=a_{0}\zeta\left(\frac{3}{2}\right)(2mT)^{3/2}+a_{1/2}(2mT)\log\left(\frac{T}{\lambda_{0}-\mu}\right).
\end{equation}
This equation is the same as the one obtained for flat space in \cite{tk} using a different asymptotic expansion of $N$. At this point following \cite{tk} one may use (\ref{no}) to replace  $T/(\lambda_{0}-\mu)$ by $N_{0}$ in the above equation
 \begin{equation}\label{ne3}
  N_{e}=a_{0}\zeta\left(\frac{3}{2}\right)(2mT)^{3/2}+a_{1/2}(2mT)\log N_{0}.
\end{equation}
However the replacement of $T/(\lambda_{0}-\mu)$ by $N_{0}$ is problematic if one considers the second term in $N_{d}$. The problem is that the term $R_{+}\widetilde{n}(s=1)$ (the functional zeta function term) contributes a $\log (\lambda_{0}-\mu)$ term which exactly cancels the $(\lambda_{0}-\mu)$ factor inside the logarithm in (\ref{ne2}). The details of this calculation which is based on scaling properties of the spectral $\zeta$ function are given in Appendix B and the corrected version of (\ref{ne3}) is found to be (\ref{EQNA}):

\begin{eqnarray}\label{EQN}
  N_{e}&=&\zeta\left(\frac{3}{2}\right)b_{0}(2mTV_{\gamma}^{2/3})^{3/2}+b_{1/2}(2mTV_{\gamma}^{2/3})\log\left(2mTV_{\gamma}^{2/3}\right)+\nonumber\\
  &&+ [\gamma b_{1/2}+R_{+}\widetilde{f}(1,\mu)](2mTV_{\gamma}^{2/3}).
\end{eqnarray}

where $\widetilde{f}(s,\mu)$ is defined in (\ref{fsm}) as

\begin{equation} \label{tildefsmu}
 \widetilde{f}(s,\mu)= \int_{0}^{\infty}dx\,x^{s-1}\,e^{-x 2m V_{\gamma}^{2/d}(\lambda_{0}-\mu)} Tr'\,e^{-x 2m V_{\gamma}^{2/d} (L-\lambda_{0})}
\end{equation}

Note that near the critical temperature, as $\mu \rightarrow \lambda_{0}$, from Appendix B we have

\begin{equation}\label{tfsmu}
   \widetilde{f}(s,\mu)= \widetilde{f}(s,\lambda_{0})+O(V_{\gamma}^{-1}).
\end{equation}

Defining $y=2mTV_{\gamma}^{2/3}$, $A=\zeta\left(3/2\right)b_{0}$, $B=b_{1/2}$, $C=
\gamma b_{1/2}+R_{+}\widetilde{f}(1,\mu)$ we write (\ref{EQN}) as $N_{e}=Ay^{3/2}+By\log y+Cy$ which upon the scaling $y=xN_e^{2/3}$ takes the form
\begin{equation}
  Ax^{3/2}-2 B(\epsilon \log \epsilon)x+C \epsilon x =1
\end{equation}
Here $\epsilon=N_e^{-1/3}$. Now we can solve this by using an expansion of the form $x=x_{0}+(\epsilon \log \epsilon)x_{1}+\epsilon x_{2}+\ldots$, where $\epsilon$ and $\epsilon\log \epsilon$ are to be treated as completely independent perturbation parameters \cite{Boyd}. As the result we obtain
\begin{equation}
  x_{0}=\frac{1}{A^{2/3}},\;\;\;\;\;\;x_{1}=\frac{4}{3}\frac{B}{A^{4/3}},\;\;\;\;\;x_{2}=-\frac{2}{3}BA^{-4/3}(\log A^{-2/3}+B^{-1}C)
\end{equation}
Second heat kernel coefficient is given as \cite{Vass, Gilkey}
\begin{equation}
  a_{1/2}=\frac{\eta}{16\pi}A_{\gamma},
\end{equation}
where $A_{\gamma}$ is the surface area of our shell in the optical metric, $\eta=-1$ for Dirichlet boundary conditions and $\eta=1$ for Robin boundary conditions.

Thus the explicit expression for $T$ is
\begin{eqnarray}\label{tcb}
  T&=&\frac{2\pi}{m}\left(\frac{n_{e}}{\zeta\left(\frac{3}{2}\right)}\right)^{2/3}\left(\frac{V_{prop}}{V_{\gamma}}\right)^{2/3}\left[1-\left(\frac{\log N_{e}^{1/3}}{N_{e}^{1/3}}\right)\frac{1}{3\zeta^{2/3}\left(\frac{3}{2}\right)}\frac{\eta A_{\gamma}}{V^{2/3}_{\gamma}}\right.\nonumber\\
  &&\left.-\left(\frac{1}{N_{e}^{1/3}}\right)\frac{1}{6\zeta^{2/3}\left(\frac{3}{2}\right)}\left(\frac{\eta A_{\gamma}}{V_{\gamma}^{2/3}}\log\frac{ 4\pi e^{\gamma}}{\zeta^{2/3}\left(\frac{3}{2}\right)}+16\pi R_{+}\widetilde{f}(1,\lambda_{0})\right)\right].\nonumber\\
\end{eqnarray}
Here we used (\ref{fsmu}) to replace $R_{+}\widetilde{f}(1,\mu)$ by $R_{+}\widetilde{f}(1,\lambda_{0})$ neglecting $O(V_{\gamma}^{-1})$ correction against the $O(V_{\gamma}^{0})$ term $ A_{\gamma}/V_{\gamma}^{2/3}$. In this expression the terms multiplied by the factor $A_{\gamma}/V_{\gamma}^{2/3}$ represent boundary corrections on the critical temperature. On the other hand the term $R_{+}\widetilde{f}(1,\lambda_{0})$ in general depends on both the bulk and the boundary data.

Let us examine (\ref{tcb}) in the absence of gravity in flat space where it is simplified by $V_{\gamma}=V_{prop}=V$, $A_{\gamma}=A_{prop}=A$. Let us assume that our box $B$ is specified by a single length scale $\ell$ with $V\propto \ell^{3}$ and $A\propto \ell^{2}$ so that the thermodynamic limit is taken by $\ell\rightarrow \infty$ while keeping $N/\ell^{3}$ fixed. Because of this assumption $A/V^{2/3}$ is scale independent. So as   $N_{e}\sim N \rightarrow \infty$ the boundary terms proportional to $A/V^{2/3}$ vanish in the thermodynamic limit. Can the term $R_{+}\widetilde{f}(1,\lambda_{0})$ have an effect on the critical temperature in the thermodynamic limit? The answer is no. In the absence of an external potential, dimensional analysis indicates that the spectrum of $L-\lambda_{0}$ is proportional to $\ell^{-2}$ and therefore the spectrum of $V^{2/3}(L-\lambda_{0})$ is scale independent. Therefore by (\ref{fsm}) $\widetilde{f}(1,\lambda_{0})$ is also scale independent and has no effect on the temperature in the thermodynamic limit. Thus in the thermodynamic limit we get the expected result $T_{c}=(2\pi/m)\left(n_{e}/\zeta\left(3/2\right)\right)^{2/3}$.

On the other hand, when the thermodynamic limit is taken in the presence of gravity it is not clear whether the correction term  $\widetilde{f}(1,\lambda_{0})$ in (\ref{tcb}), which was just shown to vanish in flat space, will vanish also in curved space or not. The relation between $\widetilde{f}(1,\lambda_{0})$, length scale $\ell$ of the box and the length scale $r_{s}$ in our metric should be worked out to check whether gravity has any effect on the critical temperature in the thermodynamic limit. This question will be addressed in a future work.

Finally let us note that for $d=2$ we get the leading order contribution
\begin{equation}
  N_{e}=b_{0}(2mTV_{\gamma})\log(2mTV_{\gamma})=a_{0}2mT\log(2mTV_{\gamma}).
\end{equation}
Thus
\begin{equation}
  n_{e}=\frac{2mT}{4\pi}\log(2mTV_{\gamma}).
\end{equation}
In the thermodynamic limit the right hand side diverges and this is the usual indication that in two dimensions condensation does not occur.

\subsection{Ultrarelativistic Case}

Let us briefly discuss the ultrarelativistic regime $\beta m << 1$ ($c=1$).  The fully relativistic result (\ref{freesum}) is in the form of the free energy in an ultrastatic spacetime and in this form its ultrarelativsitic expansion is well understood \cite{Dowker1,Actor0,Actor,Kirsten2,T2,BECUS}. Here we just quote it for $d=3$. and refer the reader to the references for its various derivations.
\begin{eqnarray}
\mathcal{F}&=&\frac{-16}{\sqrt{\pi}} \zeta(4) m^{4} a_{0} (\beta m)^{-4}  -4 \zeta(3) m^{3} a_{1/2} (\beta m)^{-3} \nonumber\\
  && -\frac{4m^{2}}{\sqrt{\pi}}\zeta(2) \left[ (2\mu^{2}-m^{2})a_{0}+a_{1}\right] (\beta m)^{-2}-\ldots
\end{eqnarray}
Thus the leading term in the expansion of the net charge $Q$ is given by
\begin{equation}
  Q=-\frac{\partial\mathcal{F}}{\partial \mu}=\frac{4}{\sqrt{\pi}}\zeta(2)  4\mu a_{0} T^{2}= \frac{4}{\sqrt{\pi}}\zeta(2)\frac{4\mu}{(4\pi)^{3/2}}V_{\gamma}T^{2}.
\end{equation}
So for the charge density we get
\begin{equation}\label{q}
  q=\frac{Q}{V_{prop}}=\frac{V_{\gamma}}{V_{prop}}\frac{4}{\sqrt{\pi}}\zeta(2)\frac{4\mu}{(4\pi)^{3/2}}T^{2}.
\end{equation}
Solving (\ref{q}) for T we get
\begin{equation}
  T = q^{1/2}\frac{\pi}{\sqrt{2\mu\zeta(2)}}\left(\frac{V_{prop}}{V_{\gamma}}\right)^{1/2}.
\end{equation}
The critical value of the chemical potential is $\mu_{c}=\epsilon_{0}$. As in the flat case the necessary condition
for the condensation is that near $T=T_{c}$
\begin{equation}
  \mu-\mu_{c}=O\left(\frac{1}{V_{prop}}\right)=O\left(\frac{1}{r_{2}^{3}}\right).
\end{equation}
Thus we arrive at
\begin{equation}\label{tasyrel}
  T\sim q^{1/2}\frac{\pi}{\sqrt{2\mu_c\zeta(2)}}\left[1-\frac{9}{8}\frac{r_{s}}{r_{2}}+\ldots \right].
\end{equation}
This reduces to the Minkowski space result in the limit $r_{s}/r_{2}\rightarrow 0$ \cite{Haber}.

\section{Conclusion}

In this work we devised a method of taking the nonrelativistic limit of the grand canonical statistical mechanics without changing the background static geometry of the spacetime. In the special case of ultrastatic spacetime we explicitly checked that our method reproduces the usual nonrelativistic grand canonical theory based on the nonrelativistic Hamiltonian given by the Schr\"{o}dinger operator on the space manifold.

In the general case of static spacetime we derived an asymptotic expansion of the resulting nonrelativistic theory which is valid near the critical temperature of a Bose system and applied the result to the study of the gravitational and boundary effects on Bose-Einstein condensation in a finite region. We have benefited from the fact that thermodynamics of a quantum field on a static spacetime can be examined by making a conformal transformation from static metric to optical metric which is an ultrastatic metric. Mellin transform and heat kernel techniques are used to make asymptotic analysis of the system. Density, temperature and pressure relations are derived using these techniques and correction terms for the critical temperature due to gravitational and boundary effects are obtained.

In this paper, all the calculations are done assuming that the system is confined away from the horizon. Near the horizon thermodynamic variables are expected to diverge. Systems near horizon will be studied in future works based on the present paper. On the other hand it is also possible to extend our analysis to the case of a self interacting nonrelativistic Bose system. Such a generalization, which we believe would be beneficial in the context of nonrelativistic scalar dark matter models \cite{Lopez, Matos,Sikivie1,Sikivie2}, will be investigated in future works as well.

\section*{Acknowledgements}
This work is supported by Bo\u{g}azi\c{c}i University BAP Grant 12501. We would like to thank A. Aliev, C. Sa\c{c}l{\i}o\u{g}lu, T. Rador and O. T. Turgut for useful conversations.

\appendix
\section*{Appendix A}

In this appendix we will derive the formulas (\ref{freesum}) and (\ref{integral}).
Starting with
\begin{equation}\label{logdet}
  \log\textrm{det}O=-\zeta_{O}'(0),
\end{equation}
\begin{equation}\label{zeta2}
\zeta_{O}(s)=\frac {1}{\Gamma(s)}\int_{0}^{\infty}dy\,y^{s-1}\textrm{Tr}\,e^{-yO}
\end{equation}
we get
\begin{equation}\label{ara1}
  \sum_{n}\log\textrm{det}(\omega_{n}^{2}+(c\sqrt{A}\pm \mu)^{2})=-\frac{d}{ds}\left[
  \frac {1}{\Gamma(s)}\int_{0}^{\infty}dy\,y^{s-1} \sum_{n}e^{-y\omega_{n}^{2}}\textrm{Tr}\,e^{-y(c\sqrt{A}\pm \mu)}\right]_{s=0}.
\end{equation}
Using the Poisson summation formula
\begin{equation}\label{poisson}
  \sum_{n}e^{- 4\pi^{2}\frac{s}{\beta^{2}}n^{2}}=\frac{\beta}{\sqrt{4\pi s}}\sum_{n}e^{- \frac{\beta^{2}n^{2}}{4s}},
\end{equation}
we get
\begin{equation}\label{ara2}
  \frac{1}{2\beta}\sum_{n}\log\textrm{det}(\omega_{n}^{2}+(c\sqrt{A}\pm \mu)^{2})=-\frac{1}{2\sqrt{4\pi}}\sum_{n}
\frac{d}{ds}\left[
  \frac {1}{\Gamma(s)}\int_{0}^{\infty}dy\,y^{s-\frac{3}{2}} e^{-\frac{\beta^{2}n^{2}}{4y}}\textrm{Tr}\,e^{-y(c\sqrt{A}\pm \mu)}\right]_{s=0}
\end{equation}
Since the $y$ integral is convergent and analytic around $s=0$ we can easily take the $s$ derivative at $s=0$ and obtain
\begin{equation}\label{ara3}
  \frac{1}{2\beta}\sum_{n}\log\textrm{det}(\omega_{n}^{2}+(c\sqrt{A}\pm \mu)^{2})=-\sum_{n}\int_{0}^{\infty}\frac{dy}{y^{3/2}}\frac{1}{2\sqrt{4\pi}}
  e^{-\frac{\beta^{2}n^{2}}{4y}}\textrm{Tr}\,e^{-y(c\sqrt{A}\pm \mu)^{2}}.
\end{equation}
The $n=0$ term in this expression is the only term that survives the $T\rightarrow 0$ limit
\begin{eqnarray}
  \mathcal{F}(T=0)&=&-\int_{0}^{\infty}\frac{dy}{y^{3/2}}\frac{1}{2\sqrt{4\pi}}
  \left[\textrm{Tr}\,e^{-y(c\sqrt{A}- \mu)^{2}}+\textrm{Tr}\,e^{-y(c\sqrt{A}+ \mu)^{2}}\right]\nonumber\\
  &=&\frac{1}{2}\left[\textrm{Tr}\,(c\sqrt{A}- \mu)+\textrm{Tr}\,(c\sqrt{A}+ \mu)\right]\nonumber\\
  &=&\textrm{Tr}\,(c\sqrt{A})=\sum_{\sigma}\epsilon_{\sigma}.
\end{eqnarray}
Here we used the zeta function identity for a positive operator $O$
\begin{equation}
  Tr O = \zeta_{O}\left(-\frac{1}{2}\right)=\frac{1}{\Gamma\left(-\frac{1}{2}\right)}\int_{0}^{\infty}\frac{dy}{y^{3/2}}\textrm{Tr}\,e^{-y O^{2}}.
\end{equation}
In what follows our main concern will be the $T\neq 0$ contribution $\mathcal{F}-\mathcal{F}(T=0)$ to the free energy for which we will, by an abuse of notation, use the symbol $\mathcal{F}$.

After subtracting the zero temperature contribution and using the subordination identity
\begin{eqnarray}\label{subord}
  e^{-b\sqrt{x}} &=&
  \frac{b}{\sqrt{4\pi}}\int_{0}^{\infty}\frac{dy}{y^{3/2}}\,e^{-\frac{b^{2}}{4y}}\,e^{-yx},
\end{eqnarray}
we can write (\ref{ara3}) as
\begin{equation}\label{prechem}
 \frac{1}{\beta} \sum_{n=1}^{\infty}\log\textrm{det}(\omega_{n}^{2}+(c\sqrt{A}\pm \mu)^{2})=-\frac{1}{\beta}\sum_{n\neq 0}\frac{1}{n}\textrm{Tr}\,e^{-\beta n(c\sqrt{A}\pm \mu)}.
\end{equation}
Thus
\begin{eqnarray}\label{freesum0}
  \mathcal{F}&=& -\frac{1}{\beta}\sum_{n\neq 0}\frac{1}{n}\left[\textrm{Tr}\,e^{-\beta n(c\sqrt{A}-\mu)}+\textrm{Tr}\,e^{-\beta n(c\sqrt{A}+\mu)}\right]\\
  &=&\frac{1}{\beta}\left[\textrm{Tr}\log(1-\,e^{-\beta(c\sqrt{A}- \mu)})+\textrm{Tr}\log(1-\,e^{-\beta(c\sqrt{A}+ \mu)})\right].
\end{eqnarray}
This is the formula (\ref{freesum}).

On the other hand applying the subordination identity (\ref{subord}) in (\ref{freesum0}) we get
\begin{eqnarray}
  \mathcal{F}&=& \sum_{n=1}^{\infty}\left[(e^{n\beta\mu}+e^{-n\beta\mu})\right]\frac{c}{\sqrt{4\pi}}\int_{0}^{\infty}\frac{du}{u^{3/2}}
  e^{\frac{(n c\beta)^{2}n^{2}}{4u}}\textrm{Tr}\,e^{-uA}\nonumber\\
  &=&\sum_{n=1}^{\infty}\left[(e^{n\beta\mu}+e^{-n\beta\mu})\right]\frac{c}{\sqrt{4\pi}}\int_{0}^{\infty}\frac{du}{u^{3/2}}
  e^{-m^{2}c^{2}\left(\frac{(\beta m^{-1})^{2}n^{2}}{4u}+u\right)}\textrm{Tr}\,e^{-u(-\Delta_{\gamma}+U)}.\nonumber\\
 \end{eqnarray}
The last line is the expression given in (\ref{integral}).

\appendix
\section*{Appendix B}

In this appendix we will give the details of the scaling argument for the $\zeta$ function leading to (\ref{ne3}).

Around $s=1$ the meromorphic continuation of
\begin{equation}
\widetilde{n}(s)=\int_{0}^{\infty}dx\,x^{s-1}\,Tr'\,e^{-x\frac{L-\mu}{(\lambda_{0}-\mu)}}=\int_{0}^{\infty}dx\,x^{s-1}\,e^{-x}Tr'\,e^{-x\widetilde{L}}
\end{equation}
is represented as
\begin{equation}
  \widetilde{n}(s)=\int_{0}^{\infty}dx\,x^{s-1}\,e^{-x }\left[Tr'\,e^{-x\widetilde{L}}-\sum_{k=0}^{d}\widetilde{a}_{k/2}x^{\frac{k-d}{2}}\right]+\sum_{k=0}^{d}\widetilde{a}_{k/2}
  \Gamma\left(s-\frac{d-k}{2}\right).
\end{equation}
The residue at $s=1$ is given by
\begin{equation}
  c_{(d-2)/2}=\sum_{l=0}^{[|\frac{d-2}{2}|]}\frac{(-1)^{l}}{l!}\widetilde{a}_{\frac{n}{2}-l}.
\end{equation}
On the other hand we also have
\begin{equation}
 \widetilde{n}(s)=\Gamma(s)\zeta_{\widetilde{L}+1}(s)=\Gamma(s)\zeta_{\frac{L-\mu}{\lambda_{0}-\mu}}(s).
\end{equation}
Let us recall the scaling property of the zeta function (see \textit{e.g.} \cite{Hawkingpath})
\begin{equation}
  \zeta_{\alpha^{-1}A}(s)=\alpha^{s}\zeta_{A}(s).
\end{equation}
Let $\alpha=2m V_{\gamma}^{2/d}(\lambda_{0}-\mu)$ then
\begin{eqnarray}
  \widetilde{n}(s)=\Gamma(s)\zeta_{\alpha^{-1}2m V_{\gamma}^{2/d}(L-\mu)}(s)= \alpha^{s}\Gamma(s)\zeta_{2m V_{\gamma}^{2/d} (L-\mu)}(s).
\end{eqnarray}
Now defining
\begin{equation}
  f(x,\mu)= Tr'\,e^{-x 2m V_{\gamma}^{2/d} (L-\mu)},
\end{equation}
we have $\Gamma(s)\zeta_{2m V_{\gamma}^{2/d} (L-\mu)}(s)$ as the meromorphic extension in $s$ of
\begin{eqnarray} 
 \widetilde{f}(s,\mu)&=& \int_{0}^{\infty}dx\,x^{s-1}\, Tr'\,e^{-x 2m V_{\gamma}^{2/d} (L-\mu)}\\
 &=& \int_{0}^{\infty}dx\,x^{s-1}\,e^{-x 2m V_{\gamma}^{2/d}(\lambda_{0}-\mu)} Tr'\,e^{-x 2m V_{\gamma}^{2/d} (L-\lambda_{0})}.\label{fsm}
\end{eqnarray}
 So around $s=1$
\begin{equation}
\widetilde{n}(s)=\alpha \left(1+(\log\alpha)(s-1)+O((s-1)^{2})\right)\left[\frac{b_{(d-2)/2}}{s-1}+R_{+}\widetilde{f}(s,\mu)|_{s\rightarrow 1}\right],
\end{equation}
where $b_{j/2}$'s are the heat kernel coefficient of $2m V_{\gamma}^{2/d} (L-\mu)$.

Since
\begin{equation}
  2m V_{\gamma}^{2/d} (L-\mu)= \alpha\frac{L-\mu}{\lambda_{0}-\mu}
\end{equation}
$b_{(d-2)/2}$ is related to $c_{d/2}$ by the usual scaling rule of the heat kernel coefficients \cite{Hawkingzeta}
\begin{equation}\label{bvec}
  b_{n/2}=\alpha^{\frac{n-d}{2}}c_{n/2},
\end{equation}
and in particular $b_{(d-2)/2}=\alpha^{-1}c_{(d-2)/2}$. Also note that using (\ref{cvea}) we get
\begin{equation}\label{bvea}
  b_{n/2}=(V_{\gamma}^{2/d})^{(n-d)/2}a_{n/2}.
\end{equation}
So
\begin{eqnarray}
  R_{+}\widetilde{n}(s=1)&=&\alpha\left[\frac{c_{(d-2)/2}}{\alpha}\ln\alpha+ R_{+}\widetilde{f}(s=1,\mu)\right]\nonumber \\
  &=& 2m V_{\gamma}^{2/d}(\lambda_{0}-\mu)\left[b_{(d-2)/2}
  \log  (2m V_{\gamma}^{2/d}(\lambda_{0}-\mu))+ R_{+}\widetilde{f}(1,\mu)\right].
\end{eqnarray}
Using this in (\ref{ndouble}) and expressing $c_{j/2}$'s in terms of $b_{j/2}$'s we arrive at
\begin{equation}
  N_{d}=b_{(d-2)/2}(2mTV_{\gamma}^{2/d})\log(2mTV_{\gamma}^{2/d})+[\gamma b_{(d-2)/2}+R_{+}\widetilde{f}(1,\mu)] (2mTV_{\gamma}^{2/d}).
\end{equation}
As mentioned above $\log (\lambda_{0}-\mu)$ terms cancel in $N_{d}$.

Similarly $N_{s}$ can also be expressed in terms of $b_{j/2}$'s as
 \begin{equation}
  N_{s}=\sideset{}{'}\sum_{n=0}^{\infty}\zeta\left(\frac{d-n}{2}\right)b_{n/2}(2mTV_{\gamma}^{2/d})^{\frac{d-n}{2}}.
\end{equation}
Therefore for $d=3$ we now have
\begin{eqnarray}\label{EQNA}
  N_{e}&=&\zeta\left(\frac{3}{2}\right)b_{0}(2mTV_{\gamma}^{2/3})^{3/2}+b_{1/2}(2mTV_{\gamma}^{2/3})\log\left(2mTV_{\gamma}^{2/3}\right)+\nonumber\\
  &&+ [\gamma b_{1/2}+R_{+}\widetilde{f}(1,\mu)](2mTV_{\gamma}^{2/3}).
\end{eqnarray}

This is the corrected version of (\ref{ne3})

Note the following analyticity property of $\widetilde{f}(s,\mu)$ in $\mu$. By the Lemma 1.4 of \cite{JorLang} the meromorphic continuation (in $s$) of $\widetilde{f}(s,\mu)$ is holomorphic in $\lambda_{0}-\mu$ for $\lambda_{0}-\mu > \lambda_{1}-\lambda_{0}$ \textit{i.e.} for $\mu < \lambda_{1}$ . Thus as $\mu\rightarrow \lambda_{0}$ we have
\begin{equation}
   \widetilde{f}(s,\mu)= \widetilde{f}(s,\lambda_{0})+O(\lambda_{0}-\mu).
\end{equation}
In particular near the critical temperature
\begin{equation}\label{fsmu}
   \widetilde{f}(s,\mu)= \widetilde{f}(s,\lambda_{0})+O(V_{\gamma}^{-1}).
\end{equation}


\begin{thebibliography}{99}
\bibliographystyle{abbrv}

\bibitem{Dowker1} J. S. Dowker, G. Kennedy, J. Phys. A: Math. Gen. \textbf{11}, 895 (1978).

\bibitem{Gibbons} G. W. Gibbons and M. J. Perry, Proc. R. Soc. Lond.  \textbf{A358}, 467 (1978).

\bibitem{Actor0} A. Actor, Nucl. Phys.  \textbf{B265}, 689 (1986).

\bibitem{Actor} A. Actor, J. Phys.  \textbf{A20}, 5351 (1987).

\bibitem{Fulling} S. A. Fulling and S. N. M. Ruijenaars, Phys. Rep. \textbf{152}, 135 (1987).

\bibitem{Dowker2} J. S. Dowker, J. P. Schofield, Phys. Rev.  \textbf{D38}, 3327 (1988).

\bibitem{Dowker3} J. S. Dowker, J. P. Schofield, Nucl. Phys.  \textbf{B327}, 267 (1989).

\bibitem{Kirsten2} K. Kirsten, Class. Quantum Grav.  \textbf{8}, 2239 (1991).

\bibitem{T1} D. J. Toms, Phys. Rev. Lett. \textbf{69}, 1152 (1992).

\bibitem{T2} D. J. Toms, Phys. Rev.  \textbf{D47}, 2483 (1993).

\bibitem{BECUS}
 L. Akant, E. Ertugrul, Y. Gul and O. T. Turgut, J. Math. Phys. \textbf{56}, 073503 (2015).

\bibitem{Alwis} S. P. de Alwis and N. Ohta, Phys. Rev. \textbf{D52}, 3529 (1995).

\bibitem{kapusta} J. I. Kapusta, \emph{Finite Temperature Field Theory}, Cambridge University Press (1993).

\bibitem{Dowkermult} J. S. Dowker, \textit{	
On the relevance of the multiplicative anomaly}, hep-th/9803200 (1998).

\bibitem{Filippi} E. Elizalde, A. Filippi, L. Vanzo and S. Zerbini, Phys. Rev. \textbf{D57}, 7430 (1998).

\bibitem{Tomsmult} J. J. McKenzie-Smith and D. J. Toms, Phys. Rev. \textbf{D58}, 105001 (1998).

\bibitem{Padmanabhan} H. Padmanabhan and T. Padmanabhan, Phys.Rev. \textbf{D84}   085018, (2011); Addendum: Phys. Rev. \textbf{D90} 089908, (2014).

\bibitem{zeldovich} Y. B. Zeldovich, JETP Lett. \textbf{12}, 307 (1970); Y. B. Zeldovich, Comm. Astrophys. and Space Phys. \textbf{3}, 179, (1971); Y. B. Zeldovich, \emph{"Creation of Particles and Antiparticles in an Electric and Gravitational Field"} in \emph{"Magic without Magic: John Archibald Wheeler" }, ed. J. Klauder, W. H. Freeman (1972).


\bibitem{Barbon} G. 't Hooft, Nucl. Phys. B256, 727 (1985);
L. Susskind and J. Uglum, Phys. Rev. \textbf{D50}, 2700 (1994);
J. L. F. Barbon, Phys. Rev. \textbf{D50}, 2712 (1994);
S. P. de Alwis and N. Ohta, \emph{On the entropy of quantum fields in black hole backgrounds}, hep-th/9412027;
R. Emperan, Phys. Rev. \textbf{D51}, 5716 (1995);
S. N. Solodukhin, Phys. Rev. \textbf{D51}, 618 (1995).

  \bibitem{Lopez} L. A. Ure\~{n}a-Lopez, Phys. Rev. \textbf{D90}, 027306 (2014).

  \bibitem{Matos} T. Matos and E. Gomez, Eur. Phys. J. \textbf{D69}, 125, (2015).

\bibitem{Sikivie1} P. Sikivie, Q. Yang, Phys. Rev. Lett. \textbf{103}, 111301, (2009).

\bibitem{Sikivie2} O. Erken, P. Sikivie, H. Tam, Q. Yang, Phys. Rev. \textbf{D58}, 063520 (2012).

\bibitem{sdm} J. Lee and I. Koh, Phys. Rev. \textbf{D53}, 2236 (1996). V. Sahni and L. Wang,
Phys. Rev. \textbf{D62}, 103517 ( 2000);
T. Matos and L.A. Urena-Lopez,
Phys. Rev. \textbf{D63}, 063506 (2001);
T. Matos and L. A. Urena-Lopez,
Class. Q. Grav. \textbf{17}, L75 (2000);
M. Membrado, A. F. Pacheco and J. Sa{\~n}udo,
Phys. Rev. \textbf{A39}, 4207 (1989);
M. R. Baldeschi, G. B. Gelmini and R. Ruffini,
Phys. Lett. \textbf{B122}, 221-224 (1983).

\bibitem{Copson} E. T. Copson, \emph{Asymptotic Expansions}, Cambridge University Press (2004).

\bibitem{deWitt} B. S. de Witt, Phys. Rev. Lett. \textbf{16}, 24, (1966).

\bibitem{Cognola} G. Cognola, L. Vanzo and S. Zerbini,  Gen. Rel. Gravit. \textbf{18},971 (1986).

\bibitem{tk} D. J. Toms and K. Kirsten, Phys. Rev. \textbf{E59}, 158, (1999).

\bibitem{Hawkingzeta} S. W. Hawking, Comm. Math. Phys. \textbf{55}, 133, (1977).

\bibitem{Hawkingpath} S. W. Hawking, \emph{The Path-Integral Approach to Quantum Gravity} in \emph{General Relativity: An Einstein Centenary Survey}, eds. S. W. Hawking and W. Israel, Cambridge University Press (1979).

\bibitem{JorLang} J. Jorgenson and S. Lang,  \emph{Basic Analysis of Regularized Series and Products}, Springer-Verlag (1993).

\bibitem{Ober}
   F. Oberhettinger,
   \emph{Tables of Mellin Transforms},
   Springer-Verlag (1974).

\bibitem{Wng}
R. Wong,
   \emph{Asymptotic Approximation of Integrals},
   Academic Press (1990).

\bibitem{Flajolet2}
P. Flajolet, X. Gourdon and P. Dumas, Theoretical Computer Science
\emph{144}, 3-58 (1995)

\bibitem{Gilkeybook} P. B. Gilkey, \emph{Asymptotic Formulae in Spectral Geometry}, Chapman and Hall/CRC (2004).

\bibitem{Thorne} K. S. Thorne, R. H. Price and D. A. Macdonald,  \emph{Black Holes The Membrane Paradigm}, Yale University Press (1986).

\bibitem{Gilkey}
 T. P. Branson and P. B. Gilkey, Comm. Partial Differential Equations
\textbf{15} 245 (1990).

\bibitem{Vass}
 D. V. Vassilevich, Phys. Rept. \textbf{388}, 279 (2003).

\bibitem{Pathria} R. K. Pathria and P. D. Beale, \emph{Statistical Mechanics}, Academic Press (2011).

\bibitem{Boyd} J. P. Boyd, \emph{Solving Transcendental Equations}, SIAM (2014).

\bibitem{Haber} H. E. Haber and H. A. Weldon, Phys. Rev. Lett. \textbf{46}, 1497 (1981).

\end{thebibliography}
\end{document}